\documentclass[a4paper,12pt]{article}

\usepackage{amsmath,amsfonts,amssymb,mathtools}
\usepackage{mathbbol,mathrsfs,dsfont,bm}
\usepackage{latexsym,wasysym}
\usepackage{cancel,slashed,units}

\usepackage{hepnicenames,feynmp-auto}

\usepackage{graphicx,rotating}
\usepackage[table,xcdraw,dvipsnames]{xcolor}

\usepackage{booktabs,multirow,multicol}
\usepackage{enumerate,expdlist}

\usepackage{soul,comment,xstring}
\usepackage[normalem]{ulem}

\usepackage[nosort]{cite}
\usepackage{hyperref}

\hypersetup{
  colorlinks=true,
  bookmarksopen=true,
  bookmarksnumbered=true,
  pdfstartview=FitH,
  linkcolor=blue!60!black,
  urlcolor=red!60!black,
  citecolor=green!50!black
}

\allowdisplaybreaks

\newcommand{\mio}[1]{}
\newcommand{\gfootnote}[1]{}

\newcommand{\beq}{\begin{equation}}
\newcommand{\eeq}{\end{equation}}
\newcommand{\be}{\begin{equation}}
\newcommand{\ee}{\end{equation}}

\newcommand{\beqa}{\begin{equation}\begin{aligned}}
\newcommand{\eeqa}{\end{aligned}\end{equation}}

\newcommand{\bea}{\begin{eqnarray}}
\newcommand{\eea}{\end{eqnarray}}

\newcommand{\bpm}{\begin{pmatrix}}
\newcommand{\epm}{\end{pmatrix}}

\renewcommand{\[}{\left[}

\newcommand{\med}[1]{\langle #1\rangle}
\newcommand{\sfrac}[2]{#1/#2}

\newcommand{\SU}{\,\mathrm{SU}}

\newcommand{\eq}[1]{~\eqref{eq:#1}}

\newcommand{\hhref}[1]{%
  \href{https://arxiv.org/abs/#1}{arXiv:#1}%
}
\newcommand{\hhrefq}[1]{%
  \IfSubStr{#1}{:}
    {\href{https://inspirehep.net/search?q=#1}{InSpire:#1}}
    {\hhref{#1}}%
}
\newcommand{\hhrefqq}[1]{%
  \IfBeginWith{#1}{10.}
    {\href{https://doi.org/#1}{doi:#1}}
    {\hhrefq{#1}}%
}

\newcommand{\doi}[1]{\href{https://doi.org/#1}{[link]}}

\makeatletter

\def\art{\@ifnextchar[{\eart}{\oart}}
\def\oart#1#2#3#4#5{%
  \textrm{#1}, \textit{#2 \textbf{#3}} \textrm{(#5) #4}%
}
\def\eart[#1]#2#3#4#5#6{%
  \oart{#2}{#3}{#4}{#5}{#6} (\textit{#1})%
}

\def\article{\@ifnextchar[{\earticle}{\oarticle}}
\def\oarticle#1#2#3#4#5#6{%
  \textrm{#1}, \textit{`#6'}, \textrm{#2 #3 (#5) #4}%
}
\def\earticle[#1]#2#3#4#5#6#7{%
  \oarticle{#2}{#3}{#4}{#5}{#6}{#7} [\hhrefq{#1}]%
}

\def\hepart[#1]#2{\textrm{#2}, {\slshape #1}}
\def\heparticle[#1]#2#3{%
  #2, \textit{`#3'} [\hhrefq{#1}]%
}

\renewenvironment{thebibliography}[1]
  {%
    \begin{multicols}{2}[\section*{\refname}]%
    \@mkboth{\MakeUppercase\refname}{\MakeUppercase\refname}%
    \list{\@biblabel{\@arabic\c@enumiv}}%
      {%
        \settowidth\labelwidth{\@biblabel{#1}}%
        \leftmargin\labelwidth
        \advance\leftmargin\labelsep
        \@openbib@code
        \usecounter{enumiv}%
        \let\p@enumiv\@empty
        \renewcommand\theenumiv{\@arabic\c@enumiv}%
      }%
    \sloppy
    \clubpenalty4000
    \@clubpenalty\clubpenalty
    \widowpenalty4000
    \sfcode`\.\@m
  }
  {%
    \def\@noitemerr{%
      \@latex@warning{Empty `thebibliography' environment}%
    }%
    \endlist
    \end{multicols}%
  }

\makeatother

\newcounter{alphaequation}[equation]
\renewcommand{\thealphaequation}{%
  \theequation\hbox to 0.6em{\hfil\alph{alphaequation}\hfil}%
}

\makeatletter

\newcommand{\eqnsystem}[1]{%
  \renewcommand{\@eqnnum}{\textrm{(\thealphaequation)}}%
  \renewcommand{\@@eqncr}{%
    \let\@tempa\relax
    \ifcase\@eqcnt
      \def\@tempa{& & &}%
    \or
      \def\@tempa{& &}%
    \or
      \def\@tempa{&}%
    \fi
    \@tempa
    \if@eqnsw
      \@eqnnum\refstepcounter{alphaequation}%
    \fi
    \global\@eqnswtrue
    \global\@eqcnt=0
    \cr
  }%
  \refstepcounter{equation}%
  \let\@currentlabel\theequation
  \def\@tempb{#1}%
  \ifx\@tempb\empty\else\label{#1}\fi
  \refstepcounter{alphaequation}%
  \let\@currentlabel\thealphaequation
  \global\@eqnswtrue
  \global\@eqcnt=0
  \tabskip\@centering
  \let\\=\@eqncr
  $$\halign to \displaywidth\bgroup
    \@eqnsel\hskip\@centering
    $\displaystyle\tabskip\z@{##}$%
    &\global\@eqcnt\@ne
    \hskip2\arraycolsep\hfil${##}$\hfil
    &\global\@eqcnt\tw@
    \hskip2\arraycolsep
    $\displaystyle\tabskip\z@{##}$\hfil
    \tabskip\@centering
    &\llap{##}\tabskip\z@\cr
}

\def\endeqnsystem{%
  \@@eqncr\egroup$$\global\@ignoretrue
}

\makeatother

\newcommand{\LEFT}{\Lambda_{\rm EFT}}
\newcommand{\NA}{\mathcal N}

\begin{document}

\begin{center}
  \bigskip\bigskip
  {\bfseries\LARGE\color{red!40!black} Challenges for minimal warm inflation\\ 
  in the Standard Model and beyond}\\
  \bigskip\bigskip
  {\bfseries Anish Ghoshal$^a$, Alessandro Strumia$^b$}\\[2ex]
  {\itshape $^a$Department of Physics and Astronomy, University of Sussex, UK}\\
  {\itshape $^b$Dipartimento di Fisica, Universit\`a di Pisa, Italia}\\
  \bigskip
  {\bfseries\large\color{blue!30!black} Abstract}
  \begin{quote}\large\color{blue!30!black} 
 A minimal realisation of warm inflation, even within the Standard Model, assumes that the inflaton $\phi$ couples via the effective operator $\phi\, G\tilde G$.
We show that the SM solutions are in tension with the plasma calculation (as each sphaleron transition releases an energy much larger than the temperature) 
and with the assumed effective operator description.
Embedding $\phi$ as the Goldstone boson of a renormalizable Peccei-Quinn theory requires an anomaly coefficient $\gtrsim100$, implying a Landau pole for QCD.
These tensions are relaxed for more strongly coupled dark gauge sectors.
  \end{quote}
\end{center}
\thispagestyle{empty}

\tableofcontents\newpage

\section{Introduction}
Warm inflation~\cite{Berera} aims to make inflation compatible with steeper potentials, smaller tensor-to-scalar ratios, and a smooth transition to radiation domination by allowing the inflaton to dissipate energy continuously into a thermal bath.
However, the couplings required to generate friction generally spoil the flatness of the inflationary potential.
A notable exception, known as `minimal warm inflation' or `Standard Model warm inflation', consists of an inflaton $\phi$ coupled to an SU($N_c$) gauge theory through the axion-like effective Lagrangian operator
\begin{equation}
-\frac{\alpha}{8\pi}\frac{\phi}{f}\,G^a_{\mu\nu}\tilde G^{a\mu\nu},
\label{eq:L}
\end{equation}
where $\alpha$ is the SU($N_c$) gauge coupling, $G^a_{\mu\nu}$ its field strength and $\tilde G^{a\mu\nu}=\epsilon^{\mu\nu\rho\sigma}G^a_{\rho\sigma}/2$.
The shift symmetry protects the {flatness of the potential}, while sphaleron transitions in the hot plasma (thermal jumps of the gauge field between vacua with different topological Chern-Simons number) produce friction
on the axion~\cite{BGK,LP}. 
The quantum chromodynamics (QCD) $\SU(3)$ part of the Standard Model (SM) gauge group may suffice~\cite{Fermions,BDZ,RR}, and several studies have attempted to identify the inflaton $\phi$ with the QCD axion~\cite{HeavyQCD,Vanilla}.

In section~\ref{sec:general}, we derive relations that are independent of the inflationary potential and constrain the parameters of the theory.
As a result:
a)  the needed temperature $T$ during inflation is not far below the cut-off $\LEFT =\sfrac{8\pi f}{\alpha}$ inferred from the coefficient of the effective operator in eq.\eq{L};
b) section~\ref{sec:id} shows that $T$ is large compared to $f$, motivating a study of the full Peccei-Quinn theory;
c) section~\ref{sec:mu} shows that the thermal plasma gets significantly modified, while in the proposed attempts the friction is computed in linear response.
In section~\ref{sec:concl} we conclude by discussing how problematic these findings are,
and how they are relaxed by replacing QCD with a dark gauge sector more strongly coupled than QCD.

\section{The temperature of warm inflation}\label{sec:general}
We consider a generic field $\phi$, not necessarily an axion, that feels a thermal friction $\Upsilon$.
The inflaton $\sigma$ is another generic field that moves along a trajectory in field space that forms an angle $\theta$ with $\phi$:
 $\dot\phi=\dot\sigma\cos\theta$, with $\cos\theta=1$ if $\phi$ is the inflaton. 
A first key parameter of warm inflation is the ratio between thermal friction $\Upsilon$ and Hubble friction $H=\dot a/a$ ($a$ is the scale factor), along the trajectory,
\begin{equation}
Q\equiv\frac{\Upsilon}{3H}\cos^2\theta.
\label{eq:Q}
\end{equation}
A second key parameter is $R$, the ratio of the thermal and vacuum contributions to the curvature perturbation. The measured power spectrum
of the comoving curvature perturbation $\mathcal R$, $\mathcal P_{\mathcal R}\approx2.1~10^{-9}$ receives a vacuum and a thermal contribution, both evaluated at
horizon exit~\cite{LPR}:
\begin{equation}
\mathcal P_{\mathcal R}=G(Q)\,(\mathcal P_{\rm vac}+\mathcal P_{\rm th})\equiv G(Q)\,(1+R)\,\mathcal P_{\rm vac}
=G(Q)\,(1+R)\left(\frac{H^2}{2\pi\dot\sigma}\right)^2.
\label{eq:PR}
\end{equation}
$G$ is a correction factor discussed below.
The stochastic force that accompanies the friction gives, for an effectively single-field trajectory~\cite{BMR,RR},
\begin{equation}
R\equiv \frac{\mathcal P_{\rm th}}{\mathcal P_{\rm vac}}
=2n_*+\frac{2\pi Q}{\sqrt{1+4\pi Q/3}}\,\frac{T}{H}
\label{eq:R}
\end{equation}
where $n_*$ is the occupation number of the inflaton modes at horizon exit, 
$n_*\approx T/H$ if they are thermalised. 
For $\cos\theta\approx1$ and $Q\ll1$ the friction is below $H$, so that the horizon-scale modes need not be
thermalised; for $n_*=0$ one has $R\approx 2\pi Q\,\sfrac{T}{H}$.
For $\cos\theta<1$ the projection of the friction is exact for the background, but the perturbations also
involve the orthogonal direction and possible entropy-to-curvature conversion, which the friction alone does
not determine: we restrict eq.\eq{R} to effectively single-field trajectories. 
Eq.\eq{R} is an analytic approximation, obtained by neglecting the dependence of $\Upsilon$ on $T$, which
decouples the inflaton perturbations from the radiation perturbations. The factor $G$ in eq.\eq{PR} is defined
as the ratio between the numerical spectrum of the full coupled system and this approximation. For
$\Upsilon\propto T^c$ with $c>0$ the coupling gives a growing mode, since a hotter region has larger friction and
is further heated: $G\approx1$ for $Q\lesssim0.1$, while for $Q\gtrsim1$ it grows as a power of $Q$~\cite{BMR,RR}.

\medskip

\begin{figure}[t]
\centering
$$\includegraphics[width=0.45\textwidth]{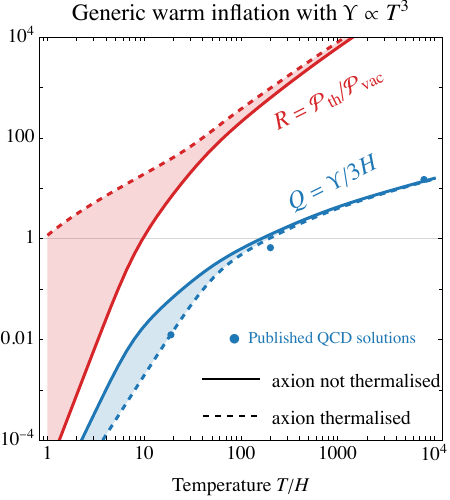}
\qquad
\includegraphics[width=0.45\textwidth]{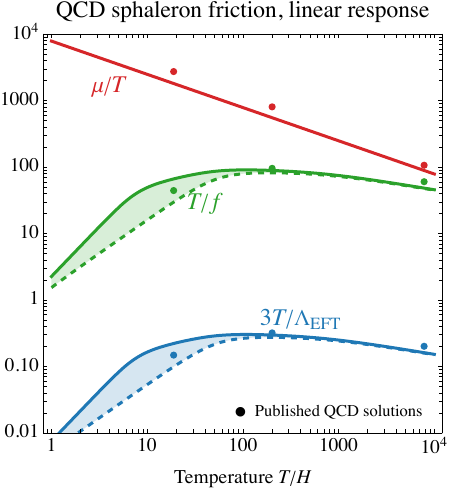}$$
\caption{\em We assume that $\phi$ is the inflaton, $\cos\theta=1$. 
\textbf{\textit{Left:}} $Q$ (blue, eq.\eq{Q}) and $R=\mathcal P_{\rm th}/\mathcal P_{\rm vac}$ (red) as functions of $T/H$,
from eq.s\eq{R} and\eq{TH} with $G(Q)$ for $\Upsilon\propto T^3$ and $g_*=106.75$, for inflaton modes not
thermalised (solid) or thermalised (dashed), with the region in between shaded. 
\textbf{\textit{Right:}} for QCD sphaleron friction in linear response, we show: 
the dimensionless free-energy bias $\mu/T$ per unit change of Chern-Simons number (eq.\eq{linearbiasscaling}); 
the ratio $T/f$ obtained from eq.\eq{identity} with the $Q$ of the left panel; 
$3T/\LEFT$ (blue), as functions of $T/H$. 
Dots: the QCD solutions of~\cite{BDZ}, ordered from left to right by increasing $Q$.}
\label{fig:Q}
\end{figure}

The temperature during warm inflation is fixed by energy balance. A stationary bath reaches
$\rho_{\rm rad} =\pi^2g_*T^4/30=\Upsilon\dot\phi^2/4H=\frac34Q\dot\sigma^2$, where $g_*$ is the effective number of relativistic energy degrees of freedom. 
Eliminating $\dot\sigma$ with eq.\eq{PR} gives
\begin{equation}
\frac{T^4}{H^4}=\frac{45\,Q}{8\pi^4g_*\,\mathcal P_{\mathcal R}} (1+R)\,G(Q).
\label{eq:TH}
\end{equation}
Under these assumptions, eq.s\eq{R} and\eq{TH} make $Q$ and $R$ functions of $T/H$ only, given $G$ and $n_*$,
independently of the inflaton potential.
The left panel of figure~\ref{fig:Q} shows $Q$ and $R$ as functions of $T/H$, using the fit {$G(Q)=1+4.981\,Q^{1.946}+0.127\,Q^{4.330}$} for $\Upsilon\propto T^3$~\cite{KMR}.\footnote{The correction $G$ is known only approximately: fitted forms~\cite{BMR,KMR} can differ from full numerical
spectra and $G$ can depend on other model parameters~\cite{BPP,SWIM}, while the thermal enhancement $2n_*$ is
debated~\cite{BPP}. Changing $G$ by a factor of 10 at fixed $T/H$ changes the $T/f$ curve in the right panel of figure~\ref{fig:Q} by about a
factor of 2, and the later eq.s\eq{muQ} and\eq{linearbiasscaling} not depend on $G$.}
For $n_*=0$ friction becomes significant, $R\gtrsim1$, for $T/H\gtrsim10$, where $Q\approx10^{-2}$. 
If instead the $\phi$ modes are thermalised ($n_*\approx T/H$) one gets $R\gtrsim1$ whenever $T\gtrsim H$. 
These relations reproduce within a factor of two the three QCD solutions of~\cite{BDZ}, 
which assume thermalised modes and use their own fit of the spectrum.
Unless stated otherwise, numerical triplets below refer, in increasing order of dissipation, to the three QCD solutions of~\cite{BDZ}, which have $Q\approx0.012$, $0.66$ and $15$.


\section{Problem 1: too large $T/f$}\label{sec:id}
We now specialise the previous general discussion to the warm inflation theory of eq.\eq{L}.
The shift symmetry forbids non-derivative couplings of $\phi$, such as $\phi^2|H|^2$ to the SM Higgs, that would give thermal corrections to the $\phi$ potential.
The $\phi G \tilde{G}$ operator approximation can hold, in the most optimistic case, at most up to energies of order 
$  \LEFT \equiv \sfrac{8\pi f}{\alpha}$.\footnote{This is an order-of-magnitude upper estimate for the validity of the local operator description, rather than a sharply defined cutoff.
A mildly higher estimate is obtained imposing partial-wave unitarity of tree-level $2\to 2$ rates~\cite{NotMinimal,2207.11262}.
We derive an improved estimate that takes into account gluon multiplicity by
imposing $\Gamma(\phi \to gg) =(N_c^2-1) \alpha^2 m_\phi^3/256\pi^3 f^2 < m_\phi$ for a massive $\phi$,
obtaining  $m_\phi < [4\pi/(N_c^2-1)]^{1/2} \LEFT \approx 1.25\LEFT$ for $N_c=3$.
Only a non-perturbative computation can perhaps provide a more precise bound.}
In a thermal plasma at weak coupling the Chern-Simons diffusion rate per unit volume is
$\Gamma_{\rm sph}=\kappa(N_c\alpha)^5T^4$ with $\kappa\sim 1$~\cite{MooreTassler}. 
The axion $\phi$ feels the friction
$\Upsilon=\Gamma_{\rm sph}/2Tf^2$, derived in linear response theory~\cite{BGK,LP}.
If this is valid one gets
\begin{equation}
Q=A\cos^2\theta\,\frac{T}{H}\left( \frac{T}{\LEFT}\right)^2,\qquad\hbox{where}\qquad
A=\frac{32\pi^2}{3}\,\kappa N_c^5\alpha^3 .
\label{eq:identity}
\end{equation}
For QCD $A\approx0.6$ neglecting light fermions, that mildly reduce $\Upsilon$ through axial-charge screening~\cite{Fermions,BDZ}.
Although sphaleron transitions are dominated by soft gluons, 
in a thermal plasma the average energy is $\med{E}=3T$ and the average centre-of-mass energy is $\med{s}^{1/2}\approx(4$-$5)T$, where $s=(p_1+p_2)^2$
is the Mandelstam invariant.
More precisely, for a Boltzmann distribution, $\med{s}^{1/2}=3\sqrt{2}T$ averaging over pairs and $\sqrt{24}T$ weighting by the collision
rate with a constant cross section; similar numerical factors arise in other cases.
Inserting in eq.\eq{identity} the $Q(T/H)$ of the left panel of figure~\ref{fig:Q} 
we obtain $3T/\LEFT$ as a function of $T/H$: this is plotted as a blue curve in the right panel.
We here assumed the most favourable case where $\phi$ is the inflaton; otherwise $T/\LEFT$ grows as $1/|\cos\theta|$.
So, achieving $R\gtrsim1$ in QCD with non-thermalised modes needs $3T/ \LEFT\gtrsim0.15$.
Furthermore, $3T/\LEFT$ reaches about $0.3$ around $Q\sim1$, and decreases at larger $Q$. 
Both values lie only moderately below the estimated limit of validity of the effective field theory (EFT).

\medskip

However, $\LEFT$ estimates the ultimate energy above which the effective operator cannot make sense.
In a full theory the effective operator can be mediated by particles lighter than $\LEFT$ and stop being valid already much below $\LEFT$.
Another relevant ratio is $T/f=8\pi T/\alpha\LEFT$. 
The green curve in the right panel of figure~\ref{fig:Q} shows that $T/f \sim 50$-$90$ along the warm-inflation solutions.
We again assumed the favourable case of $\phi$ as inflaton; in general $T/f$ increases as $1/|\cos\theta|$.
We now discuss how $f$ is related to the mass of new particles in a full theory.

\subsection{Axion-like coupling from a Peccei-Quinn anomaly}
The operator in eq.\eq{L} can arise at one loop level in a renormalizable four-dimensional theory from the anomaly of
fermions $\psi$ carrying both colour and charge $q_\psi$ under a global U(1) Peccei-Quinn (PQ) symmetry, spontaneously broken by
a complex scalar $\Phi$. 
We assume that its potential is $V=\lambda_{\rm PQ}(|\Phi|^2-v^2/2)^2$, so that $\Phi=(v/\sqrt2)\,e^{i\phi/v}$ in the vacuum.
Then the Goldstone boson  $\phi$ is the axion with decay constant 
\beq f=\frac{v}{\NA},\qquad\hbox{ where}\qquad  \NA=2 \sum_\psi q_{\psi}\, T_\psi \eeq 
is the anomaly coefficient, summed over left-handed Weyl fermions $\psi$ with PQ charges $q_{\psi}$ in units of the charge of $\Phi$, and Dynkin index $T_\psi$ of their colour representation, defined by ${\rm Tr}(T^aT^b)=T_\psi\delta^{ab}$ ($T_\psi=1/2$ for a triplet). 
For $\NA$ heavy quarks with unit PQ charge, $\NA$ is simply their number.

\smallskip

As estimated in the next section, for a non-rotating condensate in thermal equilibrium PQ restoration occurs at $T\gtrsim v$ and $\phi$ is not a Goldstone boson.
Avoiding this needs a large $\NA\gtrsim T/f$. 
For QCD, sizeable warm inflation effects need $\NA\gtrsim50$.
Standard Model quarks, as in Dine-Fischler-Srednicki-Zhitnitsky (DFSZ) axion models~\cite{DFSZ}, give $\NA=6$, and are moreover massless and screen the
friction in the electroweak-symmetric plasma~\cite{BDZ}. Heavy new quarks, as in Kim-Shifman-Vainshtein-Zakharov (KSVZ) axion models~\cite{KSVZ}, allow arbitrary $\NA$. 
But for unit
PQ charges both $\NA$ and their contribution to the running of the strong coupling are proportional to $\sum T_\psi$.
Assuming $\NA$ triplets above a common mass $M_Q$, the running QCD coupling $\alpha_3$ hits a Landau pole, where the strong coupling blows up and the theory loses perturbative control, at
\beq \Lambda_{\infty}=M_Q\exp\left[\frac{2\pi}{(2\NA/3-7)\,\alpha_3(M_Q)}\right]\approx
\begin{cases}
3~10^3M_Q & \hbox{for $\NA=50$},\\
30\,M_Q & \hbox{for $\NA=100$},
\end{cases}
\eeq 
having used $\alpha_3(M_Q)\approx0.03$.
The same holds for any fermion representations with the same $\sum T_\psi$. 
For the temperatures of~\cite{BDZ,RR} $\Lambda_\infty$ lies below the Planck mass $M_{\rm Pl}$, since $M_Q\lesssim v\sim T$.
Large PQ charges help, but require non-renormalizable couplings or small vacuum expectation values, which lower $M_Q$.
Composite axions with coloured constituents obey a similar counting. 
Axions from extra dimensions or strings need no PQ fermions, but
their decay constant is tied to the compactification or string scale,
where a whole tower of Kaluza-Klein or string modes appears.

If the axion is only a component of the inflaton, $\cos\theta<1$, eq.\eq{identity} requires an even higher
$T/f\approx(50$-$90)/|\cos\theta|$. 
Multi-field fluctuations affect $T/f$ only weakly, unless entropy-to-curvature conversion is large.
This implies that warm inflation cannot be obtained having the axion as a small component of the inflaton, 
since $\NA\gtrsim(50$-$90)/|\cos\theta|$ worsens the Landau pole.

\subsection{Warm inflation in the full Peccei-Quinn theory}
To be quantitative, we consider a renormalizable PQ theory with $\NA$ heavy colour triplets with a common Yukawa coupling
$y\,\Phi^*Q_LQ_R+\hbox{h.c.}$ in Weyl notation, so that $M_Q=yv/\sqrt2$. Since $T_3=T_{\bar3}=1/2$ and
$q_{Q_L}+q_{Q_R}=1$, each triplet contributes one unit to the anomaly coefficient.
Two regimes are possible. 

If the quarks are heavy, $M_Q\gg T$, they decouple from the plasma in the broken phase and do not screen the friction.
However, the PQ-symmetric phase, where they are massless, gains their thermal pressure $7\pi^2\NA T^4/60$.
The broken phase remains preferred only if its vacuum energy $\lambda_{\rm PQ}v^4/4$ exceeds this pressure, which requires
$\NA\gtrsim(7\pi^2/15\lambda_{\rm PQ})^{1/3}(T/f)^{4/3}\approx300$-$700$ for $\lambda_{\rm PQ}=1$. 

If the quarks are light, $M_Q\lesssim T$, for our assumed $V=\lambda_{\rm PQ}(|\Phi|^2-v^2/2)^2$ the thermal mass of the radial mode of $\Phi$ at the origin is
\beq m_\Phi^2(T)\approx-\lambda_{\rm PQ} v^2+\left(\frac{\lambda_{\rm PQ}}{3}+\frac{\NA y^2}{4}\right)T^2,\eeq
so that the multiplicity needed to enhance the anomaly also enhances the thermal correction.
The PQ symmetry must remain broken, such that $\phi$ is a Goldstone boson. 
Within the quadratic high-temperature approximation this requires
\begin{equation}
T<T_{\rm PQ}=\frac{v}{\sqrt{1/3+\NA y^2/4\lambda_{\rm PQ}}},\qquad\hbox{i.e.}\qquad
y^2<\frac{4\lambda_{\rm PQ}}{\NA}\left(\frac{v^2}{T^2}-\frac13\right).
\label{eq:Tc}
\end{equation}
In view of $\NA=v/f$ the minimal allowed $\NA$ corresponds to $v\approx T$, so that eq.\eq{Tc} requires small Yukawa couplings.
The $Q_{L,R}$ quarks are then light enough to populate the plasma and cannot be integrated out to obtain the effective operator in eq.\eq{L}.
The Goldstone potential is still protected, but friction now needs an indirect mechanism:
Yukawa interactions transfer PQ charge and energy from the rotating condensate to the quarks,
while QCD sphalerons relax the accumulated fermionic charge, allowing continued dissipation.
Sphalerons relax the common axial charge of the quarks at the rate
\beq \Gamma_A=\frac{6\NA}{N_c}\frac{\Gamma_{\rm sph}}{T^3}=2\NA\,\kappa(N_c\alpha_3)^5T,\eeq
enhanced by the multiplicity, since each transition creates chirality in all flavours~\cite{Fermions,BDZ}.\footnote{Each
transition changes the axial charge by $2\NA$ and is biased by $2\NA\mu_A$, so that
$\dot n_A=-2\NA^2\Gamma_{\rm sph}\mu_A/T$, with $n_A=\chi_A\mu_A$ and $\chi_A=N_c\NA T^2/3$.}
Within a stationary linear-response treatment, recovering an unsuppressed sphaleron drive requires the quark charge
to track the rotating condensate. This needs charge transfer faster than the combined relaxation by sphalerons and
Hubble dilution, $\Gamma_{\rm flip}\gtrsim\Gamma_A+3H$; otherwise the sphaleron drive is suppressed by incomplete
equilibration with the condensate.
The transfer proceeds through chirality flips (processes that convert left- into right-handed quarks), by mass insertions in gauge scatterings~\cite{Fermions} and directly by
the Yukawa vertex in scatterings such as $Qg\to Q\Phi$~\cite{BodekerSchroder}. The two flip rates are comparable at $T\sim v$.
Approximating the effective transfer rate as $\Gamma_{\rm flip}\approx c_{\rm flip}\alpha_3y^2T$ with $c_{\rm flip}\sim1$,
together with eq.\eq{Tc} this gives the indicative window
\begin{equation}
\frac{\Gamma_A+3H}{c_{\rm flip}\alpha_3T}\lesssim y^2<
\frac{4\lambda_{\rm PQ}}{\NA}\left(\frac{v^2}{T^2}-\frac13\right).
\label{eq:MQwindow}
\end{equation}
For $\NA=50$ and $v\approx T$, evaluating this condition on the three solutions of~\cite{BDZ} requires respectively
$\lambda_{\rm PQ}\gtrsim100$, $10$ and $0.5$, in units of $1/c_{\rm flip}$.
These estimates test the recovery of the standard sphaleron response: outside this window the quark charge cannot follow the condensate and screens the friction.
Furthermore, below $T_{\rm PQ}$ the condensate radius is already reduced,
$v(T)^2=v^2-(1/3+\NA y^2/4\lambda_{\rm PQ})T^2$, lowering $M_Q$ and $f=v(T)/\NA$: near restoration $T/f$ is even larger.
The next section discusses centrifugal force corrections.

\smallskip

The condition $T\lesssim T_{\rm PQ}$ must hold during all of inflation, in particular at its end, where $T$ typically peaks. 
Otherwise the PQ symmetry is restored, the axion and its friction disappear, and the subsequent PQ
breaking produces strings bounded by $N_{\rm DW}=\NA\gtrsim50$ domain walls.
These are cosmologically problematic, unless explicit PQ breaking lifts their degeneracy.
The potential $\Delta V$ discussed below provides such a breaking.
Alternatively, PQ restoration can be avoided by complicating the model, adding negative quartic couplings of $\Phi$ to many extra
scalars~\cite{WeinbergThermal,NonRest}.

\medskip

The inflationary potential, on which our results do not depend, is another problematic aspect of minimal warm inflation.
The QCD solutions of~\cite{BDZ} add a PQ-breaking potential $\Delta V=\lambda_\phi\phi^4$ with $\lambda_\phi\lesssim10^{-13}$. 
Such a small $\lambda_\phi$ is technically natural, being protected by the shift symmetry, 
but, since $\lambda_\phi f^4\gg\Lambda_{\rm QCD}^4$, it spoils the axion solution to the strong CP problem, unless its minimum is accurately aligned with the QCD one.
Furthermore, this potential needs to be non-periodic over a Planckian field range, many orders of magnitude larger than $f$. 
Justifying it requires monodromy~\cite{KaloperSorbo}
or a hierarchy between the decay constants of the potential and of the anomalous coupling, for which no
model is known~\cite{Vanilla,NotMinimal}.
As shown above, this issue cannot be bypassed by assuming that the inflaton is mainly another field $\sigma$,
since a small $\cos\theta$ requires an even larger $T/f$.
A different option is hybrid inflation~\cite{BGK}: the vacuum energy is supplied by a waterfall field held at
its false minimum, while only $\phi$ rolls over a small field range. This avoids a large non-periodic potential
without reducing $\cos\theta$, but the coupling between $\phi$ and the waterfall field breaks the shift symmetry.

\section{Problem 2: too large $\mu/T$?}\label{sec:mu}
The coupling\eq{L} reads $-q \phi/f$, where $q\equiv \alpha G ^a_{\mu\nu}\tilde G^{a\mu\nu}/8\pi$ is the
topological charge density, normalised so that its integral over space-time is the change of Chern-Simons
number $n_{\rm CS}$. The homogeneous axion therefore obeys
\begin{equation}
\ddot\phi+3H\dot\phi+V'(\phi)=-\frac{\langle q\rangle}{f},
\label{eq:eomq}
\end{equation}
where the average is taken over the plasma state, while the fluctuations of $q$ around it act as a
stochastic force. For $\dot\phi=0$ the plasma has $\langle q\rangle=0$. A rolling axion lowers the free energy by
$\mu\equiv\dot\phi/f$ for each unit of Chern-Simons number changed in one direction, so that
$\dot n_{\rm CS}\equiv\langle q\rangle$ becomes non-zero.
The force $F\equiv \sfrac{\langle q\rangle}{f}=\dot n_{\rm CS}/f$ opposes the axion velocity and drains its energy with power
$F\dot\phi=\mu\,\dot n_{\rm CS}$, which heats the plasma. For a stationary bath with negligible charge energy,
$4H\rho_{\rm rad} =\mu\,\dot n_{\rm CS}=3HQ_{\rm bg}\dot\sigma^2$, where $Q_{\rm bg}\equiv F\dot\phi/3H\dot\sigma^2$ is defined by the energy actually lost by the axion,
without assuming that $F$ is proportional to $\dot\phi$.
In linear response $Q_{\rm bg}=Q$ as defined in eq.\eq{Q}, but the two can differ beyond it.
Using $\dot\phi=\dot\sigma\cos\theta$ one finds
\begin{equation}
\frac{|\mu|}{T}=|\cos\theta|\,\frac{T}{f}\sqrt{\frac{2\pi^2g_*}{45Q_{\rm bg}}}.
\label{eq:muQ}
\end{equation}
This identity needs neither an inflationary potential nor a response law or perturbation spectrum.
Applied to the QCD solutions of~\cite{BDZ}, it gives $|\mu|/T\approx2700$, $800$ and $106$ for
 $Q\approx0.012$, $0.66$ and $15$, respectively.
We now show that large values of $|\mu|/T$ are required more generally.

\medskip

Eq.\eq{muQ} holds for any response of the plasma. The standard computation assumes the response for small $\mu$.
So we here denote as $\Gamma_0=\kappa(N_c\alpha)^5T^4$ the equilibrium diffusion rate, previously denoted as $\Gamma_{\rm sph}$.
A standard computation then gives\footnote{Chern-Simons number in a volume $\mathcal V$ performs a random walk with
$\langle(\Delta N_{\rm CS})^2\rangle=\Gamma_0\mathcal Vt$, i.e.\ a diffusion constant $D=\Gamma_0\mathcal V/2$. The coupling lowers the free
energy by $\mu$ per unit of $N_{\rm CS}$, giving the formal zero-flux weight $P\propto e^{\mu N_{\rm CS}/T}$. A
Fokker-Planck evolution with drift $v$ has this distribution as a zero-flux solution only for $v=D\mu/T$
(Einstein relation), giving $\langle\dot N_{\rm CS}\rangle=\Gamma_0\mathcal V\mu/2T$. Equivalently, the Kubo formula relates the
friction to the low-frequency spectral function of $q$~\cite{LP}. The result is linear as long as $\mu$ does
not modify the dynamics of the plasma itself; since the rate is odd under $\mu\to-\mu$, corrections are
relative ${\cal O}(\mu^2/\mu_*^2)$, where $\mu_*$ is the value at which the plasma dynamics changes: $\mu_*\sim T$ for
transitions activated over a large barrier, larger for the diffusive transitions of a hot unbroken gauge
theory (see below).}
\begin{equation}
\dot n_{\rm CS}=\frac{\Gamma_0\,\mu}{2T}\left[1+{\cal O}\left(\frac{\mu^2}{\mu_*^2}\right)\right].
\label{eq:linearrate}
\end{equation}
This is the friction $\Upsilon_0=\Gamma_0/2Tf^2$ of section~\ref{sec:id}, in the absence of screening. 
So eq.\eq{identity} reads $Q=\kappa(N_c\alpha)^5\cos^2\theta\,(T/f)^2(T/H)/6$, and $Q_{\rm bg}=Q$. 
Substituting it into eq.\eq{muQ}, the dependence on $f$ and $\theta$ cancels, leaving
\begin{equation}
\frac{|\mu|}{T}=
\sqrt{\frac{4\pi^2g_*}{15\kappa(N_c\alpha)^5}\,\frac HT}.
\label{eq:linearbiasscaling}
\end{equation}
Eq.\eq{linearbiasscaling} shows that, within linear
response, $\mu/T$ is fixed by $T/H$ and by the gauge theory alone, independently of $\theta$.
The red curve in figure~\ref{fig:Q} shows its value for QCD.
A general bound follows because the bath must carry a small fraction $\epsilon \equiv \rho_{\rm rad}/3\bar M_{\rm Pl}^2H^2$ of the inflationary energy, where $\bar M_{\rm Pl}=(8\pi G_{\rm N})^{-1/2}$ is the reduced Planck mass.
Then
eq.\eq{linearbiasscaling} implies
\begin{equation}
\frac{|\mu|}{T}=\sqrt{\frac{4\pi^3g_*^{3/2}}{15\sqrt{90\epsilon}\,\kappa(N_c\alpha)^5}\frac{T}{\bar M_{\rm Pl}}},
\end{equation}
which depends only weakly on $\epsilon$.
At weak coupling a high inflationary temperature necessarily comes with a large $|\mu|/T$. 
At inflationary temperatures the QCD coupling is $\alpha_3\approx0.03$, so $|\mu|/T\gtrsim\sqrt{T/10^9\,{\rm GeV}}$ for $\epsilon\lesssim10^{-2}$. 
Gauge theories with larger $\alpha$ can relax this.

\medskip

The difficulty is that classical Yang-Mills simulations have tested linear response only up to
$|\mu|/T=6$~\cite{MooreBias}. The much larger values above assume linear response in a regime where it is untested.

Linear response also requires the plasma to respond faster than the expansion rate $H$, which redshifts the soft modes: $H\ll\alpha^2T$,
the relaxation rate of the soft magnetic gluons that mediate sphaleron transitions~\cite{CSW}. 
The solutions of~\cite{BDZ} have $H/\alpha_3^2T\approx70$, $6$ and $0.2$: only the most dissipative example satisfies the condition.
Perturbative contributions to the friction, from Landau damping and plasmon decay, are instead negligible, being suppressed by slow roll~\cite{CSW}.

\subsection{Can the plasma screen $\mu$?}
The previous discussion treated the plasma as a passive bath. However, the plasma can respond by developing chemical potentials that oppose the drive.
In pure Yang-Mills, at leading order in $\alpha$ the coupling\eq{L} acts on gluons as a chemical potential for their helicity, which is not conserved:
an opposite helicity chemical potential could suppress the friction while the thermal fluctuations persist~\cite{BHM}.
Classical lattice simulations with a chemical potential for Chern-Simons number, however, find a steady drift linear in $\mu$~\cite{MooreBias}, see also~\cite{BFLS}. Within a linearized transport closure it reduces $Q$ by an order-one factor on the strongly
dissipative branch, while the relevant affinities lie outside linear response~\cite{2609.35756}.

In QCD, the main case of interest, light quarks provide a computable example. 
The anomaly  $\partial_\mu J_5^\mu=-2N_fq$ implies that each unit increase of Chern-Simons
number changes the axial charge by $-2N_f$.
Unless chirality flips remove it, the charge builds
up an axial chemical potential $\mu_A$ (taken common to all $N_f$ flavours, with $n_A\equiv n_R-n_L$ and $\mu_R=-\mu_L=\mu_A$ for the quark chiralities), which opposes the axion: topology
change is driven by $\mu_{\rm eff}=\mu+2N_f\mu_A$, with $\mu_A$ of opposite sign to $\mu$. We
check whether this screening can make $\mu_{\rm eff}/T$ small, first for the solutions of~\cite{BDZ} and then in general.
In the linear stationary treatment, screening reduces the friction to $\Upsilon_{\rm eff}=s\Upsilon_0$, with $\mu_{\rm eff}=s\mu$~\cite{Fermions,BDZ}:
$s=1$ means no screening. The solutions of~\cite{BDZ} have
$s\approx0.92$-$0.9999$, computed with the five effectively massless flavours adopted in~\cite{BDZ}, 
so screening is negligible and their $\mu_{\rm eff}/T$ remains large, while $|\mu_A|/T$ reaches about $0.9$ in the most strongly dissipative benchmark.

\smallskip

A sharper statement follows if expansion is the only sink of axial charge. 
Assume stationary $T$ and $\mu_A$, ideal massless thermodynamics, a common axial potential, and that the axion supplies all the bath energy, so that,
in terms of $\tilde\mu_A\equiv\mu_A/T$,
\begin{eqnsystem}{sys:xx}
n_A&=&\frac{N_cN_fT^3}{3}\,\tilde\mu_A\left(1+\frac{\tilde\mu_A^2}{\pi^2}\right) = -\frac{2 N_f \dot{n}_{\rm CS}}{3H},\\
\rho_{\rm rad}&=&T^4\left[\frac{\pi^2g_*}{30}
+N_cN_f\left(\frac{\tilde\mu_A^2}{2}+\frac{\tilde\mu_A^4}{4\pi^2}\right)\right] = \frac{\mu \dot{n}_{\rm CS}}{4H}.
\end{eqnsystem}
Eliminating $\dot n_{\rm CS}$ gives
\begin{equation}
-\tilde\mu_{\rm eff}\,\tilde\mu_A\left(1+\frac{\tilde\mu_A^2}{\pi^2}\right)=\frac{4\pi^2g_*}{15N_c}+2N_f\tilde\mu_A^2,\qquad\hbox{where}\qquad \tilde\mu_{\rm eff}\equiv\frac{\mu_{\rm eff}}{T}.
\label{eq:chargeconstraint}
\end{equation}
In QCD the constant term is $\approx 94$, while the left side is at most 1.1 if both chemical potentials are below $T$: 
so $\mu_{\rm eff}$ or $|\mu_A|$ exceeds about $10\,T$.
This conclusion does not rely on linear response: it holds for any dependence of the sphaleron rate $\dot n_{\rm CS}$ on $\mu_{\rm eff}$, since this rate cancels between the two balance equations.
This happens because each transition deposits both energy $\mu$ and axial charge $2N_f$,
while expansion dilutes them at comparable rates,
so it cannot independently remove the accumulated charge while preserving the heat.
Eq.\eq{chargeconstraint} neglects chirality-changing reactions: flips by flavour-dependent quark masses and Yukawa couplings provide additional sinks of axial charge, which evade it but only reduce the screening (see~\cite{Fermions,BDZ}).


\subsection{How far can linear response extend?}
We now estimate the largest value $\mu_*$ of $\mu$ for which the linear-response approximation may remain valid.

One might naively expect $\mu_*\sim T$, because detailed balance fixes the ratio of forward
and backward transition rates to $e^{\mu/T}$. Writing them as $\Gamma(\mu) e^{\pm\mu/2T}/2$, the net rate is
$\dot n_{\rm CS}=\Gamma(\mu)\sinh(\mu/2T)$.
The rate would grow exponentially for $\mu\gg T$ if $\Gamma(\mu)$ stayed equal to $\Gamma_0$.
This is what happens for transitions activated over a barrier much higher than $T$, as in the broken electroweak phase. 
It is not the case in the hot symmetric phase, where Chern-Simons number is freely changed by thermal effects,
giving rise to a diffusive dynamics.
The chemical potential then acts as a force on a diffusing variable and produces a drift linear in $\mu$ if its mobility is independent of the bias.
One then can expect that linear response can hold until $\mu$ modifies
the dynamics of the soft gauge modes themselves.

\smallskip

We then consider a different effect:
in a vacuum background the coupling in eq.\eq{L} renders tachyonic the transverse gauge modes at small momentum $k<k_{\rm ax} \equiv \alpha|\mu|/2\pi$~\cite{GFC,AnberSorbo}.
In the relevant plasma background, soft magnetic modes get suppressed at the non-perturbative scale $m_{\rm mag}=4\pi  \alpha c_{\rm mag} N_cT$.
SU(3)  lattice simulations give $c_{\rm mag}\approx0.17$-$0.19$~\cite{NakamuraMag}, close to the self-consistent estimate of~\cite{AlexanianNair}.
The value of $c_{\rm mag}$ depends on how the screening mass is defined, and $m_{\rm mag}$ should be regarded as a characteristic infrared scale rather than as a pole mass.
The ratio between $k_{\rm ax}$ and $m_{\rm mag}$,
\begin{equation}
\frac{k_{\rm ax}}{m_{\rm mag}}=\frac{|\mu|}{8\pi^2N_cc_{\rm mag}T},
\label{eq:instability}
\end{equation}
indicates when the axion starts to drive the gauge field rather than merely biasing its thermal diffusion.
Combining these effects in an approximate dispersion relation $\omega^2=k^2+m_{\rm mag}^2-k\,k_{\rm ax}$, 
the instability exists at $k_{\rm ax}>2m_{\rm mag}$ for $|\mu|>\mu_*\equiv16\pi^2N_cc_{\rm mag}T\approx(80$-$90)\,T$ in QCD.
This value is only indicative, since the soft magnetic response is dissipative and non-Abelian~\cite{1402.4174,1512.02374,hep-ph/9801430,1302.2125}, but
the solutions of~\cite{BDZ} need a larger $|\mu|/T\approx100$-$2700$.
A more solid but numerically weaker argument arises comparing $k_{\rm ax}$ to the typical thermal momentum $T$,
where physics is simpler. The solutions of~\cite{BDZ} have a dangerously large $k_{\rm ax}/T=\alpha_3|\mu|/2\pi T\sim0.5$-$10$.

Perhaps this means that axion warm inflation can still work, but in a different dynamical regime,
where the inflaton drives growing gauge fields that heat the plasma,
with thermal screening limiting the instability~\cite{FN,FMT}.
Establishing what happens needs a dedicated thermal simulation at $\mu/T\sim10^{2\hbox{-}3}$.
Beyond linear response the friction entering the background, $F/\dot\phi$, differs from the one entering the
perturbations, $\partial F/\partial\dot\phi$, and the noise is no longer fixed by the friction: a simulation should
determine the mean Chern-Simons drift, its response to $\mu$, and its fluctuations.
Classical lattice simulations of an axion coupled to SU(2), starting from a cold condensate, find that non-Abelian self-interactions moderate the instability, giving damping with a linear-response-like scaling~\cite{BFLS}; whether this persists in a thermal bath at $\mu/T\sim10^{2\hbox{-}3}$ remains open.

\medskip

Further effects arise in the PQ theories of section~\ref{sec:id}.
First, the ${\cal N}\gg1 $ fermions add degrees of freedom, $g_*\approx106.75+10.5\NA$ if they are light, making $\mu/T$ even larger.
Second, the rotation of the PQ field with angular velocity $\omega=\mu/\NA$ 
adds a centrifugal force that displaces $\Phi$ from its minimum if $\omega$ exceeds the radial mass $m_r$,
\beq
\omega\gtrsim m_r=\sqrt{2\lambda_{\rm PQ}}\, v\qquad\hbox{i.e.}\qquad\NA^2\lesssim\frac{\mu/f}{\sqrt{2\lambda_{\rm PQ}}}.\eeq
For the solutions of~\cite{BDZ} this happens for $\NA$ below about $70$–$300$ at $\lambda_{\rm PQ}=1$.
A significant centrifugal force increases  $f \gg   v/\NA $ and thereby suppresses $Q\propto1/f^2$, further excluding theories with a low number $\NA$
of colored fermions.
For the same reason, rotation can prevent thermal restoration of the PQ symmetry only at the price of a larger $f$.

\section{Conclusions and discussion}\label{sec:concl}
We showed that ``minimal warm inflation'' induced by an axion $\phi$ coupled to the QCD gluons as $\phi G \tilde{G}$ 
needs a temperature so high that:
\begin{enumerate}[a)]
\item the average thermal energy is just below the ultimate cut-off of the effective operator, $3T /\LEFT \gtrsim 0.1$;
\item  the theory applies the effective operator $\phi G \tilde{G}$ at $T/f\approx50$-$90$, which is beyond its plausible regime of validity, 
assuming that the anomaly is mediated by one loop exchange of fermions.
We computed a renormalizable Peccei-Quinn extension with $\NA$  heavy coloured fermions, finding that $\NA\gtrsim100$ are needed to enhance the effective operator,
at the price of a sub-Planckian Landau pole for the strong coupling.

\item the chemical potential induced by $\dot\phi$ is $\sim 10^{2\hbox{-}3}$ times larger than the temperature,
so that computations done in linear-response approximation, tested by simulations only up to $|\mu|/T=6$, need to be reconsidered.
In this regime plasma instabilities likely appear, possibly leading to a different form of friction.
A dedicated lattice computation is needed.
\end{enumerate}
These problems get worse if, to obtain a potential that can support inflation,
$\phi$ is only a subleading component of the inflaton, since $T/f$ grows as $1/|\cos\theta|$.
Without this possibility, inflation needs a $V(\phi)$ potential that breaks the PQ symmetry
thereby spoiling the axion solution to the strong charge-parity (CP) problem, unless its minimum is accurately aligned with the QCD minimum.

\medskip

A much larger $T/H$ would reduce $\mu/T$ and $T/f$, but in QCD, light-quark screening is then no longer diluted by expansion~\cite{BDZ}, 
and the spectral index becomes too red~\cite{RR}.
Coupling the axion to the electroweak gauge bosons as $\phi W\tilde W$, instead of to gluons, does not help.
The electroweak sphaleron rate, $\Gamma_{\rm EW}\approx18\,\alpha_2^5T^4$ where $\alpha_2$ is the SU(2)$_L$ coupling~\cite{1404.3565}, is about ten times smaller than the QCD one, so $\mu/T$ is about three times larger.
Screening by baryon plus lepton number $B+L$, which electroweak sphalerons violate, obeys a constraint analogous to that of section~\ref{sec:mu}.
A rolling axion also sources baryon number~\cite{Cohen:1987vi}, as in spontaneous
baryogenesis during warm inflation~\cite{hep-ph/0301270,1110.3971}, but at the large $\mu/T$ found here such
estimates are not under control.

A possible way forward is to abandon minimality and replace QCD SU(3) with a hypothetical dark SU($N_c$) gauge sector.
Its gauge coupling $\alpha$ at inflationary temperatures can be larger than $\alpha_3$: for example $N_c=3$ and $\alpha=0.1$ give $3T/\LEFT\approx0.02$-$0.04$ and $T/f\approx2$-$4$ between $R=1$ and $Q=1$.
Then a few heavy fermions $\NA\gtrsim (N_c\alpha)^{-5/2}$ suffice, and the theory can be asymptotically free, $11N_c>2\NA$, possibly in all its couplings as in~\cite{2003.10446,2007.00005}.
One still has $\mu/T\approx25$-$100$. 
The inflationary potential can also come from a separate, e.g.\ hybrid, sector~\cite{BGK,HeavyQCD},
within the limits on decay-constant hierarchies~\cite{Vanilla,NotMinimal}.
The resulting dark glueballs must either decay sufficiently early or constitute viable dark matter. 

%


{\small
\paragraph{Acknowledgements.}
AS thanks Anthropic for a `Claude Team plan for scientists' discount.
We thank Andrea Tesi and ChatGPT for comments, and Claude, which took an active part in this work, from calculations to writing.
AG acknowledges support from the Royal Society, UK, funding reference: NIF\ R1\ 253963.

\appendix\small

\end{document}